\documentclass[doublecol]{epl2}

\def\be{\begin{equation}}
\def\ee{\end{equation}}
\usepackage{graphicx}
\usepackage{epstopdf} 
\usepackage{dcolumn} 
\usepackage{bm}
\usepackage{subfigure}

\newcommand{\ra}{{\rightarrow}}
\newcommand{\ket}[1]{| #1 \rangle}
\newcommand{\Ham}{{\cal H}}

\begin{document}

\title{Ground state factorization and correlations with broken symmetry}

\author{Bruno Tomasello \inst{1} \and Davide Rossini  \inst{2} \and Alioscia Hamma   \inst{3} \and Luigi Amico \inst{1}}

\institute{
 \inst{1} 
MATIS-INFM-CNR $\&$ Dipartimento di Fisica e Astronomia, 95123 Catania, Italy \\
\inst{2} 
NEST, Scuola Normale Superiore $\&$ Istituto Nanoscienze-CNR, Piazza dei Cavalieri 7, I-56126 Pisa, Italy\\
\inst{3} 
Perimeter Institute for Theoretical Physics, 31 Caroline St. N, Waterloo ON, N2L 2Y5, Canada \\
}

\abstract{
  We show how the phenomenon of factorization in a quantum many body system is of collective nature. 
  To this aim we study the quantum discord $Q$ in the one dimensional $XY$ model in a transverse field.
  We analyze the behavior of $Q$ at both the critical point and at the non critical factorizing field. 
  The factorization is found to be governed by an exponential scaling law for $Q$. 
  We also address the thermal effects fanning out from the anomalies occurring at zero temperature. 
  Close to the quantum phase transition, $Q$ exhibits a finite-temperature crossover 
  with universal scaling behavior, while the factorization phenomenon results 
  in a non trivial pattern of correlations present at low temperature. 
}


\maketitle

\textit{Introduction.---} 
The concepts of symmetry and correlations pervade all the modern many body physics~\cite{anderson}. 
A system consisting in a very large number of particles can be found in different 
phases and the Landau-Ginzburg paradigm of symmetry breaking characterizes the various phases 
in terms of different symmetries. Different quantum phases are separated by Quantum Phase Transitions (QPTs), 
which are driven by tuning an external control parameter $h$ across a critical value $h_c$~\cite{sachdev}. 

Nevertheless, in the past twenty years it has been understood that symmetry cannot explain 
quite all the phases of matter~\cite{wenbook}. Indeed, different patterns of correlations 
can define different quantum phases featuring unconventional transitions~\cite{wen_quantumphase}. 
Examples in many-body physics come from studies on high Tc superconductors, as well as intermetallic compounds 
(heavy fermions) and fractional quantum Hall liquids~\cite{wenbook, superc, coleman_heavy}.

Here we analyze quantum correlations in a many body system addressing the {\em quantum discord}, 
beyond the generic notion of ``correlations in a quantum system''~\cite{zurek,vedral}. 
Quantum correlations are not all captured by entanglement, because a non vanishing 
quantum discord results for certain separable (mixed) states~\cite{zurek}. 
This study addresses some new features of the quantum phases involved in the phenomenon 
of symmetry breaking. Besides the critical behavior of the quantum discord 
at the quantum phase transition, the discord displays dramatic changes also at a 
non critical value of the control parameter $h_f \neq h_c$, where quantum correlations vanish, 
thus producing a factorized classical state~\cite{factorization, giampaolo}. 
Such factorization can even occur within the symmetry broken phase,
and it consists in the sudden reshuffling of quantum correlations, 
leading to a transition in the entanglement pattern~\cite{fubini,amico}.
We show that this {\it correlation transition} at $h_f$ is governed by a new class of scaling laws,
thus signaling a collective nature of the phenomenon, even if it is not associated to any symmetry breaking. 
We speculate that the factorization can be associated to exotic 
quantum phase transitions that are not described by symmetry breaking 
but by a reorganization of entanglement patterns without symmetry breaking, 
like in topological quantum phase transitions~\cite{topqpt}.

We complete our study by detecting how the quantum critical and the factorization point 
affect the quantum discord at low-temperature, thus opening the way towards actual observations~\cite{Yurischev}.

\textit{Quantum Discord in the XY model.---} 
The total amount of correlations in a bipartite (mixed) 
quantum state $\hat{\rho}_{AB}$ is given by the mutual information 
$I_{AB} \equiv S(\hat{\rho}_A) + S(\hat{\rho}_B) - S(\hat{\rho}_{AB})$, 
where $S(\hat{\rho}) = - {\rm Tr} [\hat{\rho} \log_2 \hat{\rho}]$ is the von Neumann entropy. 
On the other hand, classical correlations can be defined 
in terms of the quantum conditional entropy: 
$S(\hat{\rho}_{AB} | \{\hat{B}_k\}) = \sum_k p_k S(\hat{\rho}_{AB}^{(k)})$, where 
$\hat{\rho}_{AB}^{(k)} = \frac{1}{p_k} (\hat{I} \otimes \hat{B}_k) \, \hat{\rho}_{AB} \, (\hat{I} \otimes \hat{B}_k)$ 
is the state of the composite  system $AB$, conditioned to the outcome $\hat{B}_k$ 
(being a set of projectors representing a complete measurement of the subsystem $B$) 
of the measurement, with probability $p_k = {\rm Tr} [(\hat{I} \otimes \hat{B}_k) \, \hat{\rho}_{AB} 
\, (\hat{I} \otimes \hat{B}_k)]$.
The amount of classical correlations $C$ is obtained by
finding the set of measurement on $B$ that disturbs the least the part $A$, i.e.,
by maximizing 
$C = \max_{\{\hat{B}_k\}} \big[ S(\hat{\rho}_A) - S(\hat{\rho}_{AB} | \{\hat{B}_k\}) \big]$
(here we restrict to projective measurements)~\cite{vedral,zurek}.
The quantum discord is given by: $Q=I-C$.
In a pure state, $Q$ reduces to entanglement. A mixed state though, 
may contain quantum correlations that are not accounted in the lack of separability 
(see Ref.~\cite{zurek} for examples). 

The model we study is the spin-$1/2$ chain with $XY$ exchange couplings in a transverse field $h$:
\be
   \hat{\Ham} = -\sum_j \left(
   \frac{1+\gamma}{2}\hat{\sigma}_{j}^x \hat{\sigma}_{j+1}^x +
   \frac{1-\gamma}{2} \hat{\sigma}_{j}^y \hat{\sigma}_{j+1}^y +
   h \hat{\sigma}_{j}^z\right) \, ,
   \label{eq:XYmodel}
\ee
where $\hat{\sigma}^\alpha_j$ ($\alpha = x,y,x$) are the Pauli matrices on site $j$,
$\gamma \in \left( 0, 1 \right]$ denotes the $xy$ anisotropy, 
while $h$ is the transverse magnetic field strength. 
The Hamiltonian $\hat{\Ham}$ is diagonalized by means of a
Jordan-Wigner transformation followed by a Bogoliubov rotation 
in momentum space~\cite{mcoy}.
In the range of $\gamma$ we consider hereafter, the system displays 
a continuous QPT at $h_c=1$ of the Ising universality class with critical 
indices $\nu=z=1$, $\beta=1/8$. 
Because of superselection rules, 
the region $|h|<h_c$ 
is magnetically ordered and the global $Z_2$ symmetry 
is broken in the thermodynamic limit with non vanishing order parameter
$g_x = \langle \hat{\sigma}^x \rangle$; elsewhere the system is a paramagnet.
Although the ground state of $\hat{\Ham}$ is generally entangled, 
for specific values of $\gamma$ and $h$ it is completely separable~\cite{amico}. 
Besides the trivial cases $h=0$ and $h\ra\infty$, where $\ket{\psi_{gs}}$ is fully polarized, 
there is a non trivial line of factorization $h_f^2+\gamma^2=1$ where, 
for $\langle \sigma_x \rangle \neq 0$, 
$\ket{\psi_{gs}} = \prod_j |\psi_j \rangle$~\cite{mcoy}, within the findings 
of~\cite{factorization, giampaolo}.
This line corresponds to an accidental degeneracy of the Hamiltonian~\cite{korepin,giorgidepasquale},
while the entanglement pattern swaps from parallel to anti-parallel, 
with a logarithmically divergent range 
of bipartite entanglement (``entanglement transition''~\cite{fubini, amico06}).

In order to compute the classical correlations $C_r$ and the quantum discord $Q_r$ 
of two spins $A$ and $B$ at distance $r$, one needs to access the single-spin and 
the two-spin reduced density matrices $\hat{\rho}_A$ and $\hat{\rho}_{AB}(r)$
(see, e.g., Ref.~\cite{palacios} for an explicit expression of the generic two-spin matrix
in a system with global phase flip symmetry).
Hereafter we focus on the symmetry-broken ground state and on the thermal states
of Eq.~(\ref{eq:XYmodel}).
For $Z_2$-symmetric states, the non vanishing entries of $\hat{\rho}_A$ 
and $\hat{\rho}_{AB}(r)$ can be evaluated analytically in terms of $g_z=\langle \hat{\sigma}^z \rangle$ 
and $g_{\alpha \alpha}(r) = \langle \hat{\sigma}^\alpha_j \hat{\sigma}^\alpha_{j+r} \rangle$~\cite{mcoy}.
In that case we use a fully analytic treatment for the quantum discord,
obtaining the {\it thermal ground state} as the zero temperature limit of such class of states~\cite{sarandy}. 
For symmetry-broken states, $g_{xz}(r)$ and $g_x$ also need to be accessed. 
Since the expression of $g_{xz}(r)$ is cumbersome~\cite{johnson}, 
in the latter case we resort to the numerical Density Matrix Renormalization Group (DMRG)
for finite systems with open boundaries~\cite{dmrg}.

\begin{figure}[!t]
  \centering
  \includegraphics[width=\columnwidth]{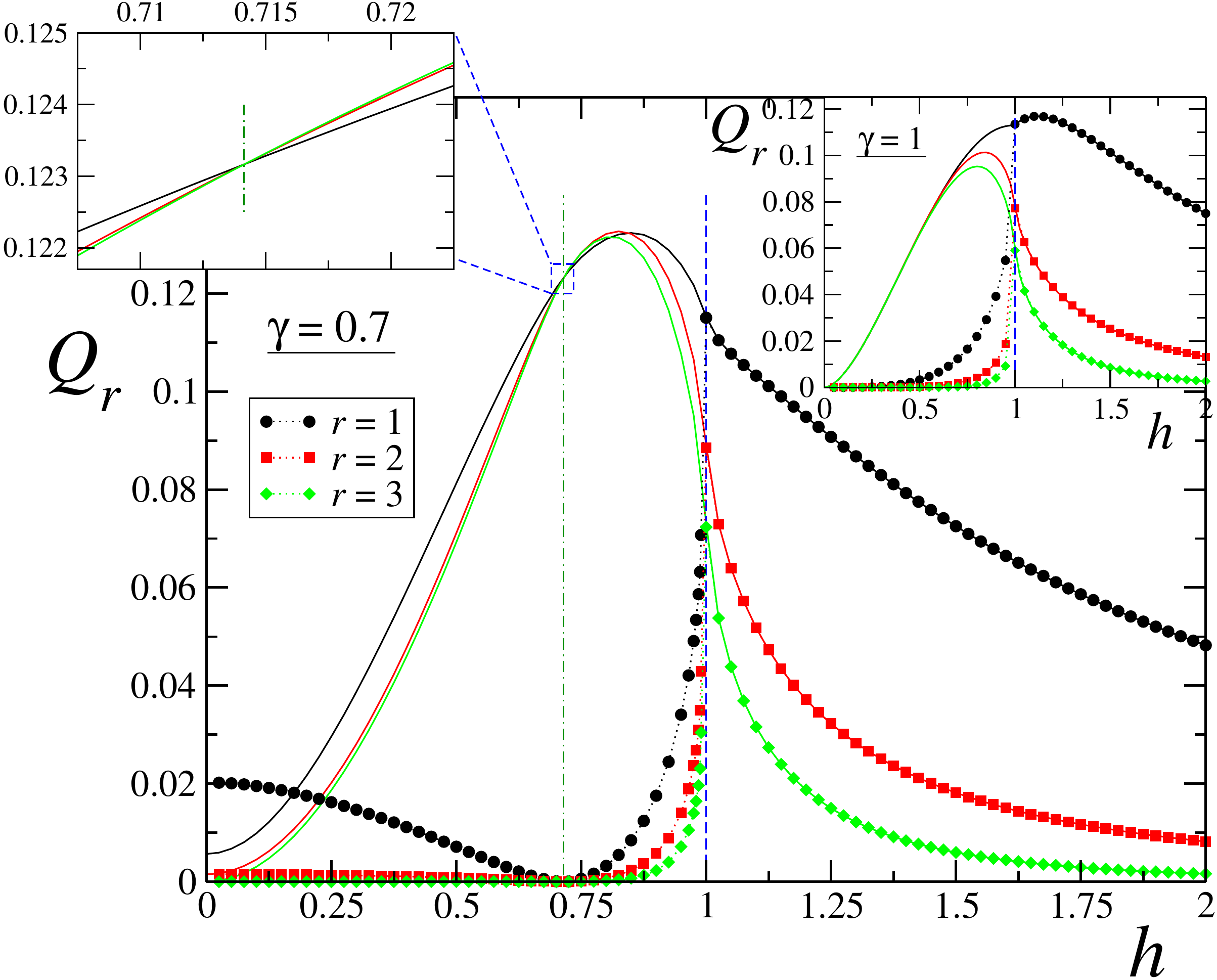}
  \caption{Quantum discord $Q_r(h)$ between two spins at distance $r$ in the $XY$ model 
    at $\gamma = 0.7$ (main plot and left inset) and $\gamma = 1$ (right inset), 
    as a function of the field $h$.
    Continuous lines are for the thermal ground state, while symbols denote the
    symmetry-broken state obtained by adding a small symmetry-breaking 
    longitudinal field $h_x = 10^{-6}$ and it was computed with DMRG
    in a chain of $L=400$ spins; simulations were performed by keeping $m=500$ states
    and evaluating correlators at the center of the open-bounded chain. 
    For $\gamma=0.7$ and at $h_f \simeq 0.714$, in the symmetric state all the curves 
    for different values of $r$ intersect, while after breaking the symmetry
    $Q_r$ is rigorously zero.
    At the critical point $Q_r$ is non analytic, thus signaling the QPT. 
    In the paramagnetic phase, there is no symmetry breaking to affect $Q_r$.}
  \label{qd}
\end{figure} 

\textit{Ground state.---} 
As displayed in Fig.~\ref{qd}, the difference between the quantum discord $Q_r$ 
for the thermal ground state and for the symmetry-broken state is always finite 
in the ordered phase (the mutual information $I$ does have the same behavior). 
Moreover, quantum correlations are typically much smaller
deep in the ordered ferromagnetic phase $h < h_c$, rather than in the paramagnetic one $h > h_c$. 
Nonetheless, as we shall see, they play a fundamental role to drive the order-disorder
transition at the QPT, where $Q_r$ exhibits a maximum, as well as the correlation
transition at $h_f$, where $Q_r$ is rigorously zero. 

\begin{figure}[!t]
  \centering
  \includegraphics[width=\columnwidth]{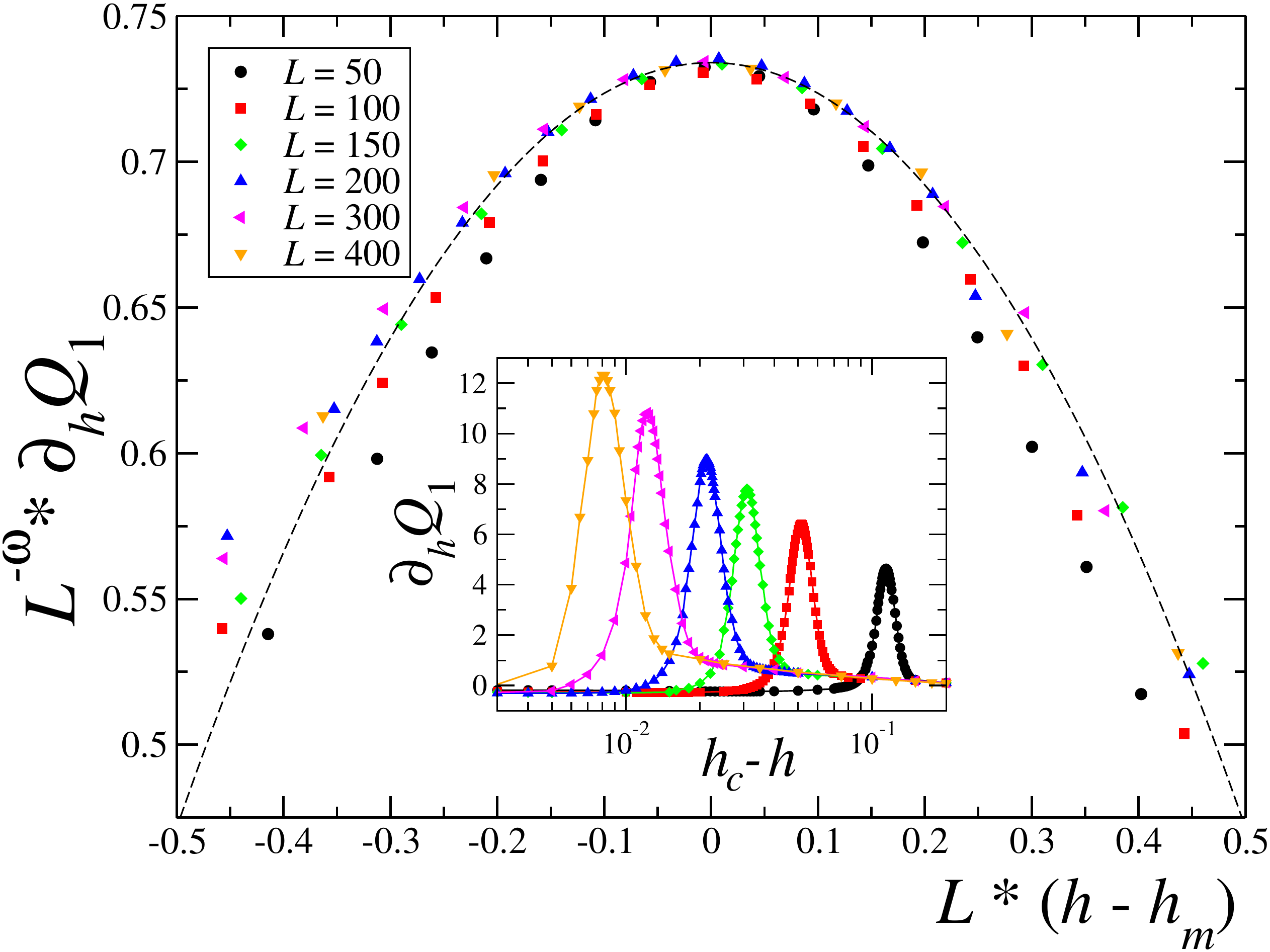}
  \caption{Finite-size scaling of $\partial_h Q_1$ for the
    symmetry-broken state in proximity of the critical point $h_c$. 
    Displayed data are for $\gamma= 0.7$. 
    The first derivative of the quantum discord is a function of $L^{-\nu}(h-h_m)$ only, 
    and satisfies the scaling ansatz $\partial_h Q_1 \sim L^\omega \times F[L^{-\nu}(h-h_m)]$,
    where 
    $h_m$ is the renormalized critical point at finite size $L$ and $\omega=0.472$. 
    We found a universal behavior $h_c - h_m \sim L^{-1.28 \pm 0.03}$ 
    with respect to $\gamma$. 
    Inset: raw data of $\partial_h Q_1$ as a function of the transverse field.}
  \label{qcritical}
\end{figure}

Let us first focus on the quantum critical point, where the QPT is marked by a divergent 
derivative of the quantum discord (see also~\cite{first_ising,sarandy,maziero}). 
Such divergence is present at every $\gamma$, for the symmetry broken state;
on the other hand, for the thermal ground state, it is not present at $\gamma=1$. 
A thorough finite-size scaling analysis is shown in Fig.~\ref{qcritical} proving that $z=\nu=1$.  
For the thermal ground state (in the thermodynamic limit), we found that $\partial_h Q_r$ 
diverge logarithmically as $\partial_h Q_r \sim  \ln |h-h_c| $, within the Ising universality class.
%

\begin{figure}[!t]
  \centering
  \includegraphics[width=\columnwidth]{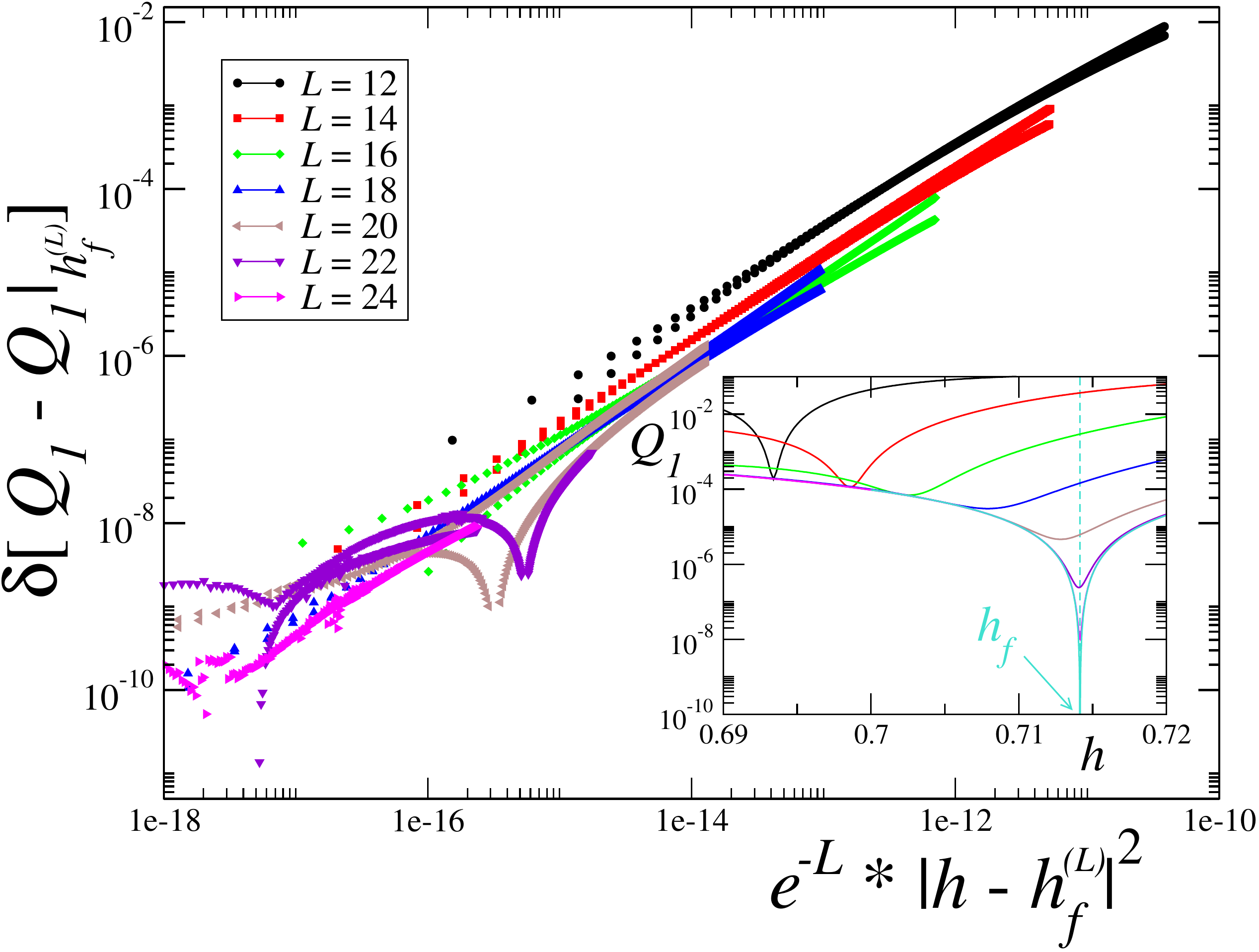}
  \caption{Scaling of $Q_1$ close to the factorizing field, for $\gamma = 0.7$:
    we found an exponential convergence to the thermodynamic limit, 
    with a universal behavior according to $e^{-\alpha L}(h-h_f^{(L)})$,
    $\alpha \approx 1$ [$h_f^{(L)}$ denotes the effective factorizing field 
    at size $L$, while $\delta(Q_1) \equiv Q_1^{(L)} - Q_1^{(L \to \infty)}$].
    Due to the extremely fast convergence to the asymptotic value,
    already at $L \sim 20$ differences with the thermodynamic limit
    are comparable with DMRG accuracy.
    Inset: raw data of $Q_1$ as a function of $h$. The cyan line is for $L = 30$
    so that, up to numerical precision, the system behaves at the thermodynamic limit.}
  \label{tscaling}
\end{figure}

At the factorizing field $h_f$, all the correlation measures are zero in the state 
with broken symmetry (see symbols in Fig.~\ref{qd}); in particular, we numerically found 
a dependence $Q_r \sim (h-h_f)^2 \times \big( \frac{1-\gamma}{1+\gamma} \big)^r$ close to it. 
Such behavior is consistent with the expression of correlation functions 
close to the factorizing line obtained in Ref.~\cite{baroni}, and here appears 
to incorporate the effect arising from the non vanishing spontaneous magnetization. 
The factorization phenomenon can be traced also for the thermal 
ground state~\cite{ciliberti}: it is the unique value of the field where 
the same quantum correlations are present at any length scale (left inset of Fig.~\ref{qd}). 
We found a rather peculiar dependence of $Q_r$ on the system size, 
converging to the asymptotic value $Q_r^{(L \to \infty)}$ with an exponential 
scaling behavior (see Fig.~\ref{tscaling}). The picture elucidated here 
suggests the existence of a non trivial mechanism leading to the factorization of the ground state. 
In~\cite{amico06,fubini}, it was shown that $h_f$ marks the transition between 
two different patterns of entanglement. The factorization is thus a new kind of zero-temperature 
transition of collective nature, not accompanied by a change of symmetry, 
and with a scaling law that is new in the panorama of the cooperative phenomena 
in quantum many-body systems. We emphasize, though, that this transition does not correspond 
-in this model- to a QPT. The factorization occurs without any non analyticity in the 
ground-state wave function $\ket{gs(h)}$ as a function of $h$, as it is shown by
the ground-state fidelity $\mathcal F(h)\equiv |\langle gs(h)|gs(h+\delta h\rangle|$. 
This quantity (which can detect both symmetry breaking and non-symmetry breaking QPTs~\cite{fidelity,topqpt}), 
is a smooth function at $h_f$. So there is no QPT here. Nevertheless, the phenomenon 
of factorization can accompany a topological QPT~\cite{topqpt}. 
We speculate that the scaling laws associated to topological QPTs are those associated 
to factorization or other phenomena of entanglement reorganization.
At the level of spectral properties of the system, we interpret this result 
as an effect of certain {\it competition between} states belonging to different parity sectors 
for finite $L$~\cite{giorgidepasquale}; as these states intersect, the 
ground-state energy density is diverging {\it for all finite $L$} (such divergence, though, 
vanishes in the thermodynamic limit). 

\textit{Finite temperature.---} 
In order to check how the observed phenomena are resilient with respect to thermal fluctuations, 
we analyze the quantum correlations at finite temperature.
The low-temperature behavior is influenced by the proximity to critical and factorizing fields.
Close to $h_c$, the physics is dictated by the interplay between thermal and quantum fluctuations 
of the order parameter.
A $V$-shaped diagram in the $h-T$ plane emerges, characterized by 
the cross-over temperature $T_{cross} = |h-h_c|^{z} $ fixing the energy scale~\cite{sachdev}. 
$T\ll T_{cross}$ identifies two semiclassical regimes. 
In the quantum critical region $T\gg T_{cross}$, quantum and thermal effects cannot be resolved; 
here the critical properties dominate the physics of the system, even at finite temperature.
Close to $h_f$ and at small $T$, the bipartite entanglement remains vanishing in a finite 
non linear cone in the $h-T$ plane~\cite{amico,amico06}. 
Thermal states, though, are not separable, and entanglement is present 
in a multipartite form~\cite{toth}. In this regime the bipartite entanglement results to be non monotonous,
and a reentrant swap between parallel and antiparallel entanglement is observed~\cite{amico06}.

At any temperature $T>0$, the state is $Z_2$-symmetric. 
By inspection of Fig.~\ref{qd}, and since $Q_r$ is a continuous function of $T$ for 
finite temperatures, we conclude that $Q_r$ is discontinuous as the temperature is switched on,
in all the phase $h < h_c$.
Such discontinuity is also observed in the entanglement, even if in that case 
it is much less pronounced and occurs only for $h < h_f$~\cite{palacios}.
We now analyze how criticality and factorization modify the fabric 
of purely quantum correlations in the $h - T$ plane. 

\begin{figure}[!t]
  \centering
  \includegraphics[width=0.9\columnwidth]{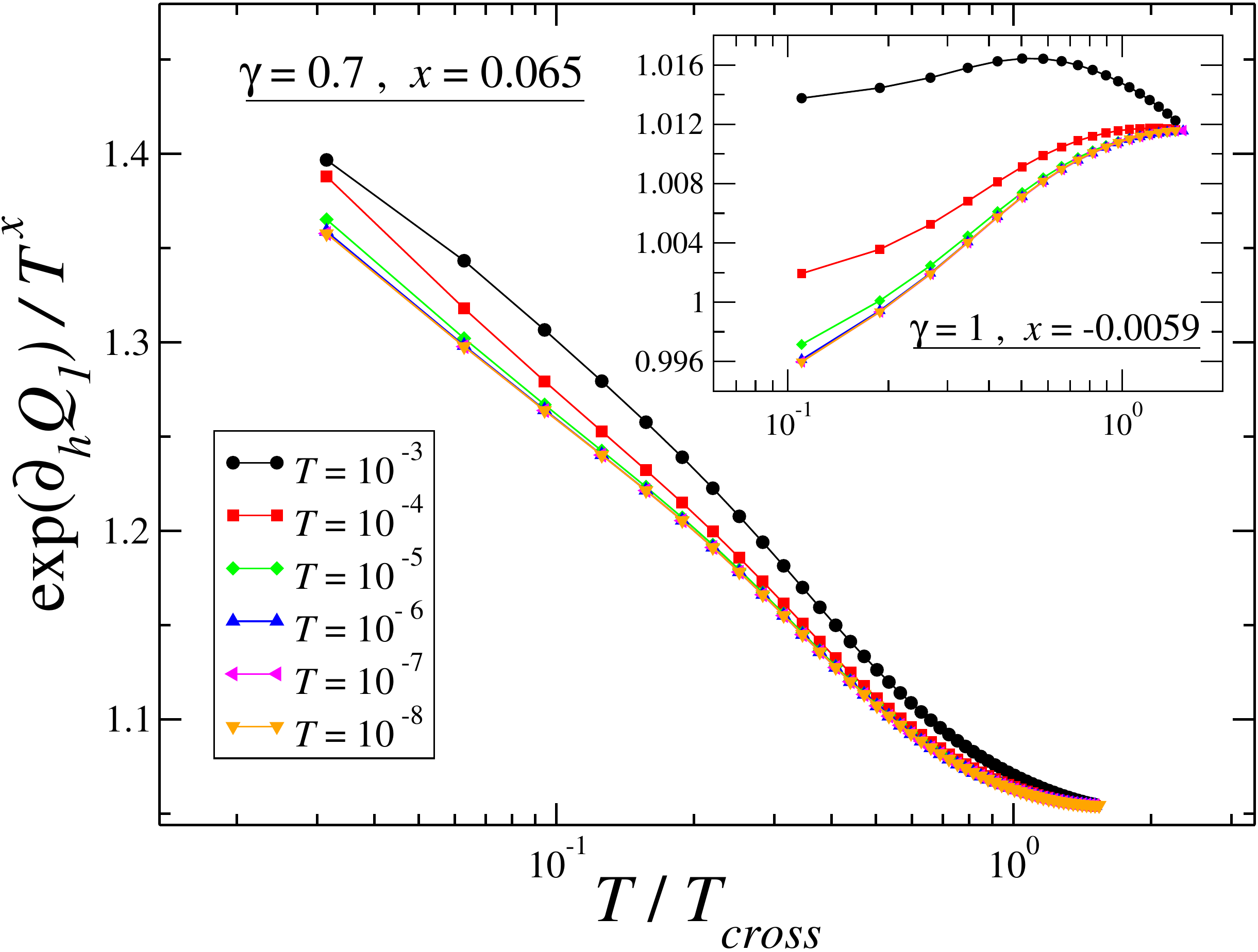}
  \caption{Finite-temperature scaling of the quantum discord for the thermal state 
    close to the critical point. 
    The logarithmic scaling is verified : along the critical line we found   
    $\partial_h Q_1 \vert_{h_c} \sim x \ln(T) + k$, with $x = 0.065$ for $\gamma = 0.7$.
    The scaling function $F$ shows a data collapse close to the critical point.
    Inset: same analysis for the Ising case ($\gamma= 1$); 
       we found an analogous scaling behavior with  $x=-0.0059$.}
  \label{scaling_T}
\end{figure} 

\begin{figure}[!t]
  \centering
  \subfigure[]{\includegraphics[width=0.24\textwidth]{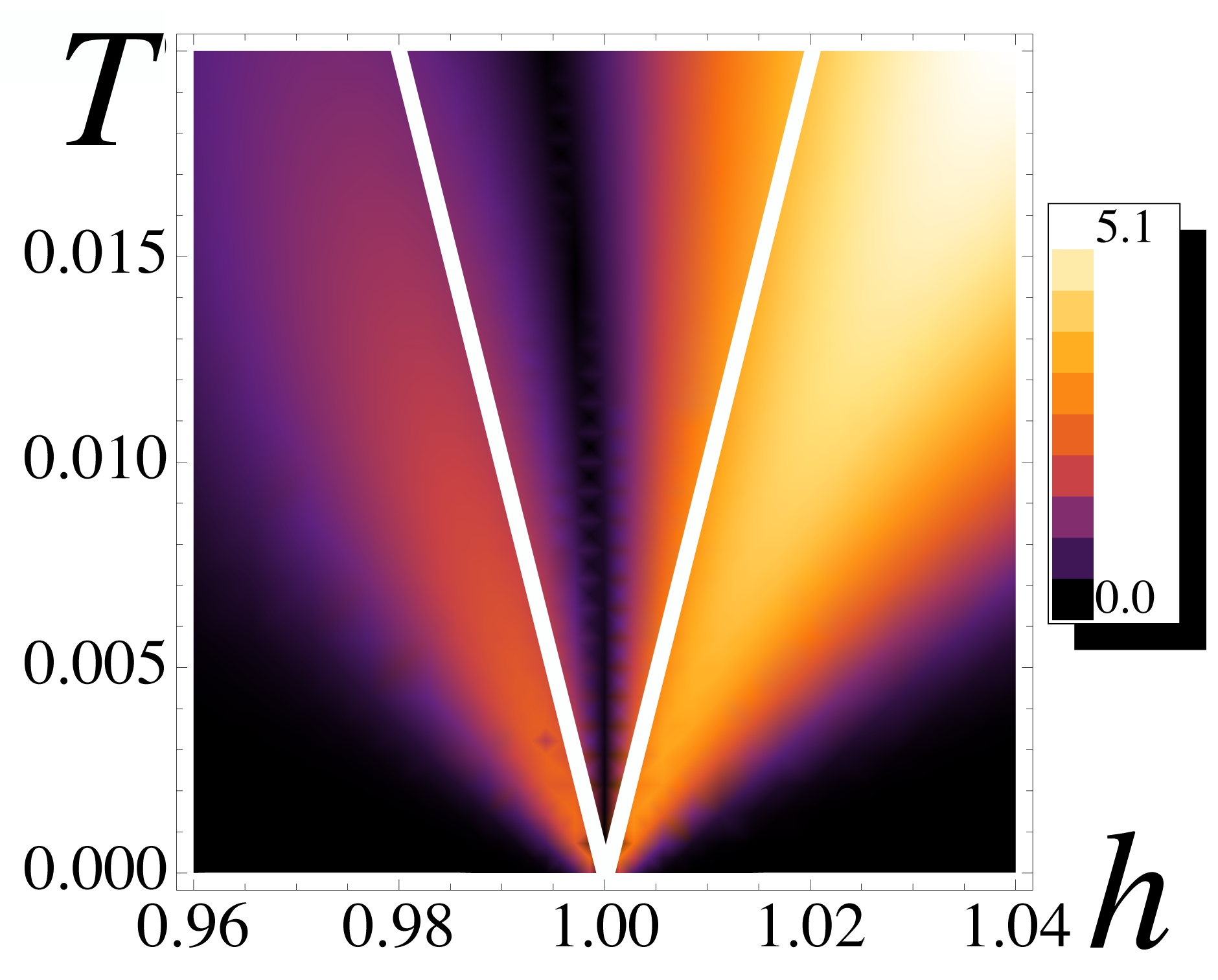}}
  \subfigure[]{\includegraphics[width=0.235\textwidth]{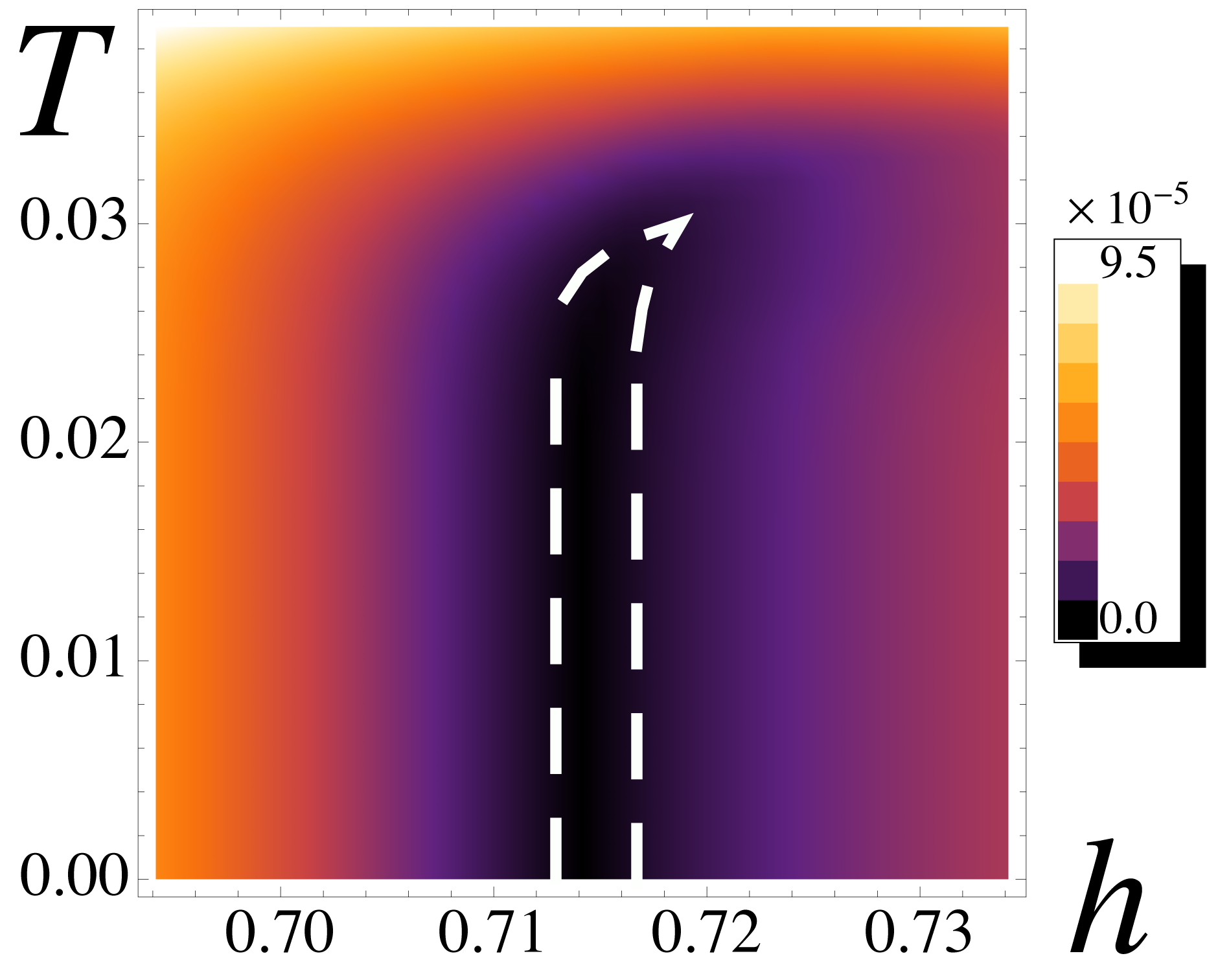}}
  \caption{a) Density plot in the $h-T$ plane of $\partial_T [Q_1/C_1]$ close to $h_c$; 
    along the critical line the ratio $Q_1/C_1$ is constant with respect to the temperature. 
    The solid straight line ($T= |h-h_c|$) marks the boundary of the quantum critical region.
    b) Average quantum discord displacement: $\overline{\Delta Q_r} =2\sum_{i,j=1}^m|Q_{r_i}(T) - Q_{r_j}(T)|/m(m-1)$
    for $m=5$ fanning out from the factorizing point $h_f \sim 0.714$,
    where all correlations coincide at any length scale $r$, as evidenced Fig.~\ref{qd}
    left inset. Dashed line is for guiding eyes. Here $\gamma=0.7$.}
  \label{discord_thermal}
\end{figure}

The first aspect we consider is the thermal scaling close to the QPT. 
The scaling ansatz $\partial_h Q_r = T^x \, F(T/T_{cross})$, where 
$T_{cross} \equiv |h-h_c|^{\nu z}$~\cite{sachdev}, is verified 
in Fig.~\ref{scaling_T} for $r=1$. 
Remarkably, the discord scales also in the Ising case $\gamma=1$, 
despite $\partial_h Q$ is not diverging at $T=0$~\cite{sarandy}.
We then discuss the interplay between classical and quantum correlations. 
In Fig.~\ref{discord_thermal}a we show
$\partial_T [Q_1/C_1]$ in the $h-T$ plane, namely the sensitivity of the relative 
strength between quantum and classical correlations to thermal fluctuations. 
We found a $V$-shaped diagram, where the ratio is constant along the critical line $h=1$ 
in the quantum critical region $T\gg|h-1|$, while it explores the largest changes 
along the crossover region. We remark the asymmetry of Fig.~\ref{discord_thermal}a 
between $\Delta < 0,\Delta > 0$, taking into account that the mechanism leading 
to the two corresponding semiclassical regimes traces back to quantum ($\Delta > 0$) 
or thermal ($\Delta< 0$) fluctuations~\cite{sachdev}.  
 
We now move to $h_f$, where, for the thermal ground state, factorization 
is marked by the fixed point in $Q_r$ (see left inset of Fig.~\ref{qd}). 
This originates a non trivial pattern of correlations: $Q_r(T) \simeq Q_{r'}(T)$ 
for any $r,r'$. We quantify this behavior by analyzing the average displacement 
between different $ Q_r $ fanning out from the fixed point in the thermal ground state at $h=h_f$ 
(see $\overline{\Delta Q_r}$ in Fig.~\ref{discord_thermal}b).

\textit{Outlook and perspectives.---} 
We studied purely quantum correlations quantified by the quantum discord $Q_r$ 
in the quantum phases involved in a symmetry-breaking QPT.
Even if $Q_r$ results relatively small in the symmetry-broken state
as compared to the thermal ground state, it underlies key features 
in driving both the order-disorder transition across the QPT at $h_c$, and the 
correlation transition across the factorizing field $h_f$.
The critical point is characterized by a non analyticity of $Q_r$ found in the Ising universality class. 
Close to $h_f$, $Q_r$ displays uniquely non trivial 
properties: in the thermal ground state quantum correlations are identical 
at all scales; for the symmetry-broken state we identified a novel exponential scaling, specific 
for the factorization phenomenon emerging as a new kind of collective phenomenon occurring 
in the ground state of the system.
We remark that this can occur without changing the symmetry of the system, 
as a signature of the fact that quantum phases and entanglement are more subtle than what 
the symmetry-breaking paradigm says. 
Although in model~Eq.(\ref{eq:XYmodel}) the factorization happens deep in the symmetry broken phase, 
its behavior is also particularly relevant in the context of QPTs involving 
topologically ordered phases~\cite{topentropy}, which are believed to occur 
because of a change of the global pattern of entanglement~\cite{topqpt} instead of symmetry. 

At finite temperatures a discontinuity of $Q_r$ with $T$ is evidenced in the
whole ordered phase $h<h_c$. We expect such discontinuity to be present 
also for models with finite $T_c$. 
We proved that $Q_r$ displays universal features, and it exhibits a crossover behavior;
in particular the quantum critical region is identified by the condition $Q(T)/C(T)=Q(0)/C(0)$
along the critical line. 
We have found that a non trivial pattern of quantum correlations fans out
from the factorization of the ground state (where $\overline{\Delta Q_r} = 0$), 
opening the way to experimental detection of the phenomenon.

We thank A. De Pasquale, R. Fazio, S. Montangero, D. Patan\'e, M. Zannetti for useful discussions.
The DMRG code released within the PwP project (www.dmrg.it)
has been used. Research at Perimeter Institute for Theoretical
Physics is supported in part by the Government of Canada through NSERC and
by the Province of Ontario through MRI.
DR acknowledges support from EU through the project SOLID.


\end{document}